\documentclass[aps,prl,showpacs,twocolumn,
nofootinbib]{revtex4}
\usepackage[dvips]{epsfig}
\usepackage{graphicx}
\usepackage{amsmath}

\begin{document}


\title{{\Large\bf  Avoiding BBN Constraints on Mirror Models for Sterile
Neutrinos }}

\author{\bf  R.N. Mohapatra and  S. Nasri }
\affiliation{ Department of Physics, University of Maryland, College Park,
MD-20742, USA}
\date{July , 2004}
\begin{abstract}
We point out that in models that explain the LSND result for
neutrino oscillation using the mirror neutrinos, the big bang
nucleosynthesis constraint can be avoided
by using the late time phase transition that only helps to mix the
active and the sterile neutrinos. We discuss the
astrophysical as well as cosmological implications of this proposal.
\end{abstract}

\vskip1.0in

\maketitle

\section{Introduction}
 The existence of neutrino oscillations has
now been confirmed for solar and atmospheric neutrinos as well as
for reactor and accelerator neutrinos. It is remarkable that all
the data from many different experiments can be well understood in
terms of only three neutrinos that mix among themselves. They
imply very narrow ranges of both the mass difference squares among
these neutrinos as well as mixings.

There is however another piece of evidence for oscillations which
if confirmed will require severe departure from the successful
three neutrino scheme just mentioned. It is the apparent
observation of the muon anti-neutrino oscillating to the electron
type anti-neutrino in
 the Los Alamos LSND\cite{lsnd} experiment. An attempt was made to confirm
this result by KARMEN\cite{karmen} collaboration which
 eliminated a large fraction of
the parameter space allowed by LSND.  It is hoped that the
Mini-BOONE experiment at FERMILAB currently under way will settle
the issue in near future\cite{louis}.

If the LSND experiment is confirmed, one straightforward  way to
understand the results would be to postulate the existence of one
or more extra neutrinos with mass in the eV range that do not
interact with the W boson, the so-called sterile neutrinos and
have them mix the known neutrinos. There have been various
versions of this suggestion depending on the detailed mass
arrangement of the sterile neutrinos with respect to the known
ones: they are known in the literature as the 2+2\cite{caldwell},
3+1,\cite{other} as well as 3+2\cite{sorel} models. Of the three,
3+1 model seems less disfavored than the 2+2 by the null results
of other oscillation experiments. However the more recently
proposed 3+2 scenario\cite{sorel} that involves two sterile
neutrinos is apparently in better agreement wil all data than the
others.

The major challenge posed by the sterile neutrino for theory is to
understand its ultra-lightness despite its being a standard model
singlet. A class of particle physics models that successfully
answer this challenge are the mirror matter models. The basic
assumption of these models is that there is an identical copy of
the standard model (both constituents and forces) in
nature\cite{volkas,bere} that co-exists with the familiar standard
model matter and forces. It is then clear that the same mechanism
that keeps the active neutrinos light, will also keep the mirror
neutrinos light which can therefore play the role of the sterile
neutrinos. These models are inspired by the superstring theories and have
been widely discussed\cite{bere,babu}. Phenomenological and astrophysical
constraints on these models have also been extensively
discussed\cite{strumia,mohan}.

Sterile neutrino models for LSND face two cosmological huddles
 that we would like to address in this paper. The issues
are: how to make them consistent with (i) our understanding of big
bang nucleosynthesis(BBN) and (ii) the recent bounds on neutrino
masses from WMAP observations. The first problem is that BBN
 allows the number neutrinos $N_\nu$, in equilibrium when the temperature
of the Universe is one MeV is restricted by $^4$He and $D_2$
observations to be  very close to three\cite{dolgov}. On the other
hand for $\nu_s$ mass in the eV range and mixing in the few per
cent range required to explain the LSND data, rapid $\nu_e-\nu_s$
oscillations would lead to $N_\nu~=~4$ for the 3+1 and 2+2
scenarios and $N_\nu~=~5$ for the 3+2 scenario.

The WMAP\cite{wmap} constraints are on the sum of all neutrino
masses in equilibrium at the epoch of structure formation which
corresponds to a temperature around an eV.
 According to \cite{hannestad},
$\sum m_\nu \leq 1.38$ eV for one sterile and $\sum m_\nu \leq
2.12$ eV for two extra ones assuming that they went into
equilibrium at the BBN epoch. Thse constraints are also quite
important since taken at face value, they would seem to rule out
the 3+2 model for LSND.

It is therefore important to look for scenarios that may allow one
to avoid both the above constraints while at the same time
providing an explanation of the LSND experiment. Recently, it has
been suggested that\cite{chacko}  by using late time phase
transition to generate the masses and mixings of both the active
and sterile neutrinos, one can avoid both these constraints. In
ref.\cite{chacko}, it is shown that this can be achieved by
endowing two  scalar fields $\phi$ with vevs in the 100 keV range
so that at the BBN time the sterile as well as the active
neutrinos are massless. As a result there is no oscillation among
them that can bring the sterile neutrinos into equilibrium. Since
the sterile neutrinos decouple from Hubble expansion at very high
temperatures, their abundance at the BBN epoch is suppressed
leading to concordance with the BBN constraints. Cosmological
signatures of generic  models of this type have been given in
Ref.\cite{chacko}.

In this paper we propose an alternative way to avoid the
cosmological constraints using the same idea of late time phase
transitions. We show that if the sterile neutrinos are the mirror
neutrinos, we need only generate the mixing between the active and
the sterile neutrino (and not masses) by the late time phase
transition to avoid the BBN and WMAP constraints. An advantage of
this model is that the contribution of the sterile neutrinos to
the energy density of the universe at the BBN epoch is giverned by
a free parameter unlike the model of ref.\cite{chacko}. We further
find a convenient realization of mirror model with the seesaw
scale in the TeV range which implies that we must employ the
double seesaw mechanism\cite{valle} to get small neutrino masses.
We construct explicit scenarios with late phase transition 
 and discuss their cosmological and astrophysical
implications. The detailed field theoretical models for them can be worked
out but we do not discuss them here.

\section{An extended mirror model}
 We start by reminding the reader about the generic features of the mirror
 models where one assumes that the universe consists not only of the
observed
standard model particles and forces but also coexisitng
 with an identical but different set of constituents
experiencing analogous but different gauge forces. Gravity is
however common to all the particles. The forces are dictated by
the the gauge group $G\otimes G$ where one of the
gauge groups $G$ acts in the standard model sector and on its
fermions whereas the other acts in the other and on the mirror
fermions. The fermion spectra on both sides are identical. Mirror
symmetry keeps the gauge couplings equal but the the effective
strength of various forces in both sectors may be different due to
different patterns of symmetry breaking.
 These models are inspired by the superstring
theories as well as M-theory inspired brane-bulk models that have
been widely discussed.

As is clear, the neutrinos in the mirror sector do not experience
the known weak interactions and will not therefore appear in the Z
and W-decays. They can therefore play the role of the sterile
neutrinos used in the interpretation of the LSND results. The
mixing between the active and mirror neutrinos can arise in a
manner consistent with gauge invariance (see below for details).
For purposes of notation, we denote all particles and parameters
of the mirror sector will by a prime over the corresponding
familiar sector symbol- e.g. mirror quarks are $u',d',s',$ etc and
mirror Higgs field as $H'_{u,d}$ etc.

Before proceeding further, let us discuss the origin of the masses
and mixings for the active and sterile neutrinos. For this
purpose, we extend
 the standard model gauge group in each sector to $SU(2)_L\times
U(1)_{I_{3R}}\times U(1)_{B-L}$, which is an anomaly free gauge
group in the presence of the right handed neutrino $\nu^c$. All
the fermions have obvious quantum numbers under the gauge group.
We add a gauge singlet chiral fermion, $S$ in each sector, one per
family. Mirror symmetry requires that we do the same in the mirror
sector. In order to get the standard model gauge group from the
extended group in each sector, we need to add a pair of new Higgs
bosons $\Delta (1,+\frac{1}{2}, -1)$ and a conjugate field
$\bar{\Delta}(1,-\frac{1}{2}, +1)$ in the visible sector and two
similar fields in the mirror sector. We then add a gauge singlet
Higgs field $\chi$ ( and a mirror $\chi'$), which gives Majorana
mass for the singlet fermions $S$ by an interaction of the form
$\lambda'_{\alpha\beta}(S_\alpha
S_{\beta}\chi+S'_{\alpha}S'_{\beta}\chi'$.

The full superpotential relevant for neutrinos in each sector can
be written as:
\begin{eqnarray}
W~=~h_\nu LH_u\nu^c + \lambda_1 \nu^c \Delta S +\lambda' SS\chi;
\end{eqnarray}
where we have omitted an identical set of terms for the mirror
sector and have suppressed the generation index. For three
generation case that we will be interested in, $h_\nu$,
$\lambda_1$ and $\lambda'$ are $3\times 3$ matrices.

 The $U(1)_{I_{3R}}\times U(1)_{B-L}$ part of the gauge symmetry is broken
by the vev of field $<\Delta>$ assumed to be in the multi-TeV
range. We choose the vev of $\chi$ field to be in the GeV range.
It is then clear that this leads to the double seesaw form
\cite{valle} for the $(\nu, \nu^c,S)$ mass matrix:
\begin{equation}
M_\nu = \left(\begin{array}{ccc}
0 & h_\nu v & 0  \\
h^T_\nu v & 0 & \lambda_1 v_R\\
0 & \lambda^T_1 v_R & \lambda'<\chi>
\end{array}\right).
\end{equation}

There is also a similar matrix for the mirror neutrinos. This
leads to the light neutrino mass matrix of the form:
\begin{eqnarray}
{\cal M}_\nu~=~ h_\nu M^{-1}_R\lambda'<\chi>{ M^{-1}_R}^Th^T_\nu
v^2
\end{eqnarray}
where $M_R~=~\lambda_1 <\Delta>$. It follows that if we choose
$<\chi>$ about a GeV, $h_\nu\sim 10^{-1}$ and $\lambda'\sim
10^{-4}$, then for $\frac{v}{M_R}\sim \frac{v'}{M'_R}\sim 10^{-2}
$, the neutrino masses are in the $0.1$ eV range as required by
observations. Also, typical neutrino mass textures can be built
into the coupling matrix $\lambda'$.

A similar situation will occur in the mirror sector, where we can
choose $<\chi'>$ about a factor of 10 higher to get $m_{\nu_s}$ in the eV
range to fit LSND (henceforth, we will call the sterile neutrinos $\nu'$
as $\nu_s$).

In order to generate mixing between the active and sterile
neutrinos, we postulate the existence of a scalar field $\phi$
that mixes the two sectors. This can only be done through an
interaction of the form $\beta SS'\phi$ \cite{brahma}. A simple tree
level diagram via the exchange of $\nu^c, S$ and ${\nu^c}',S'$
then leads to an effective coupling of the form:
$\lambda^{''}\frac{LH_uL'H'_u\phi}{M_RM'_R}$, where
$\lambda^{''}\simeq \beta h^2_\nu \sim 10^{-2}$. We assume as in
ref.\cite{chacko} that the vev of the field $<\phi>\sim 100$ keV,
so that at the BBN epoch the active and sterile neutrinos are
unmixed\footnote{ The smallness of the $\phi$ vev can be justified
if we embed our theory into a brane bulk scenario and have the
$<\phi>$ vev occur in a distant brane and get transmitted to the
our brane via a bulk scalar field.} . The resulting $\nu-\nu_s$
mixing is then given by: $m_{\nu-\nu_s}\simeq
\lambda''\frac{v^2<\phi>}{M_RM'_R} \sim 10^{-6}<\phi>$ for
$\lambda''\simeq 10^{-2}$. This gives the right order of mixing
for the LSND experiment. One difference between our model and that
of ref.\cite{chacko} is that, late phase transition was used in
\cite{chacko} to generate both masses and mixings whereas in our
case, only the mixing need to be generated at a late stage. As we
will see, the mirror model has the advantage that contribution of
the sterile neutrinos to the energy density of the Universe at the
BBN epoch is given by an arbitrary parameter. The model can
therefore work even if the BBN constraints on the number of extra
neutrino species tightened further.

\section{BBN, asymmetric inflation and neutrino mixings}
Before we address the issue of neutrino mixings and BBN, we note
that in the mirror model, we have three light neutrinos, a mirror
photon and a mirror electron that could be potential contributors
to the energy density at BBN epoch and affect the success of BBN.
In order to reduce their contribution to a negligible level, the
idea of asymmetric inflation was proposed in the second paper of
ref.\cite{bere}, according to which it is assumed that the reheat
temperature after inflation in the mirror sector is lower than
that in the visible sector by a factor of 10 or so i.e.
$T'_R\simeq T_R/10$. If the interactions linking the two sectors
are such that they are not in thermal contact for $T\leq T_R$,
then Hubble expansion will roughly maintain the ratio of the two
temperatures till the BBN epoch apart from minor corrections
arising from particle annihilation in both sectors. Thus at the
BBN epoch, the total contribution of the light mirror particles to
$\rho_{tot}$ is at the level of about $10^{-3}\rho_\gamma$ which
therefore keeps the predictions of standard BBN unchanged. This
will also keep the number densities of the light particles such as
mirror neutrinos and mirror photons suppressed.

\section{Constraints and consequences}
The first constraints on the parameters of the model come from the
fact that at temperatures below the reheating temperature after
inflation, the interaction rates for $L+H\rightarrow L'+H'+\phi$
should be out of equilibrium, otherwise, the two sectors will be
in the same thermal bath and at BBN, the effective $N_\nu$ will
far exceed the allowed limit due to the fact that the population
of the sterile neutrinos will build up their density to the level
of ordinary active neutrinos. Only exception is if the reheat
temperature after inflation is below an MeV which we do not invoke
here. This translates into the following constraints in the three
temperature regimes discussed below.

\noindent{\it (i) $T\geq <H>,<H'>$:} In this region, the condition for
being out-of-equilibrium is
\begin{eqnarray}
T_D\leq \left(g^{1/2}_*\frac{M^4_R}{M_{P\ell}}\right)^{1/3}
\end{eqnarray}
This inequality implies that for $T\gg$ TeV (note that the mirror Higgs
mass is expected to be in the TeV range), the visible and the
mirror sector in our model are in equilibrium. 
 So to be consistent with
BBN requirements, we must require that the reheat temperature
after inflation be less than about a TeV or so.

\noindent{\it (ii) $0.1~ MeV \leq T\leq <H>,<H'>$:} 
 For $T\leq M_{H'}\sim $ 1 TeV, the Higgs fields decay instantly
In this
regime, the effective interaction connecting the visible to the
mirror sector is the coupling $\nu\nu_s\phi$ with a strength given
by $g_{\nu\nu_s\phi}\sim
\lambda''\left(\frac{v}{M_R}\frac{v'}{M'_R}\right)\approx
10^{-6}$. In discussing whether this interaction is in
equilibrium, it has been noted in ref.\cite{chacko} that the process
$\nu\nu_s\to
\phi$, vanishes in the limit of $m_\phi~=~0$ by energy momentum
conservation. The rate for this process must therefore be
proportional to $m^2_\phi(T)$. This effective thermal mass is
given by $\kappa^2T^2/16\pi^2$. If we choose the
scalar self coupling $\kappa $ to be of order $10^{-2}$, due to
the fact that the number density of $\nu_s$ is down by a factor of
$10^{-3}$, we expect the rate $\Gamma(\nu\nu_s\to \phi)\simeq
10^{-21}T$. This leads to the conclusion that above $T\sim $ 100
keV, the interactions connecting the visible with the mirror
sector are out of equilibrium. Thus (i) and (ii) together then
help to satisfy the BBN constraints.

\noindent{(iii) $T\leq 0.1$ MeV:}

This regime is below the scale of $<\phi>$. Therefore,
$\nu-\nu_s\to \phi$ is kinematically forbidden since $Im\phi$
becomes a pseudo-Goldstone boson. The only process that can lead to
production of sterile neutrinos is $\nu\nu\to \nu_s\nu_s$. The
rate for this process is given by $10^{-25}T$. This process is in
equilibrium below $T\leq 100$ eV. 
 Below this temperature the mirror sector and the
standard model neutrinos will thermalize without a significant
transfer of energy. When the temeparture drops to $T\simeq m_s$ ,
the sterile neutrinos decay into one of the three active neutrinos and
the singlet Higgs and finally, when below the temperature $\sim $ 1 eV
which is the mass of the $\phi$, it will annihilate via the process
$\phi\phi\rightarrow \nu\nu$ leaving only a bath of active neutrino.

To calculate the final temperature of neutrino bath in terms of the photon
temperature, we first remember that the boson $\phi$ is part of a
supersymmetric
multiplet whose scalar field part has only one surviving imaginary part
and the real part has decoupled at very high temperature. We assume that
the imaginary part is also superlight. Then we procedd
through the following steps:

Just below $T\simeq m_e$, electron-positron annihilation heats up the
photons leading to the relation
$T^0_\nu~=~\left(\frac{4}{11}\right)^{1/3}T_\gamma$. The $\nu_s$ and
$\phi$
at this stage are not in thermal contact with the active neutrinos. Once
the temperature of the universe cools below 100 eV, 
$\nu-\nu_s\phi$ system comes into full thermal equilibrium. Using energy
conservation and taking the effect of the incomplete $\phi$ supermultiplet
into acccount, we get
\begin{eqnarray}
(3+\frac{11}{7}+n')T^4_{\nu+\phi+\nu_s}=3 {T^0_\nu}^4
\end{eqnarray}
where $n'$ is the number of sterile neutrinos  and the factor
$\frac{11}{7}$ take into account the contribution of the fermionic part of
 singlet Higgs superfield $\phi$. 
As the universe cools below the mass of $\nu_s$, the $\nu_s$ decay to
$\nu+\phi$. Using entropy conservation at this stage, we get
\begin{eqnarray}
(3+\frac{11}{7})T^3_{\nu+\phi}~=~(3 +\frac{11}{7}+n') T^3_{\nu+\phi+\nu_s}
\end{eqnarray}
Using the above two equations, we get for the
temperature of the ${\nu + \phi}$ system
\begin{equation}
T_{\nu + \phi}^4 = \frac{3}{3 + \frac{11}{7}}[1 + \frac{n'}{3 +
\frac{11}{7}}]^{1/3}(\frac{4}{11})^{4/3} T_{\gamma}^4
\end{equation}
Noting that $\rho_\nu \propto (3+\frac{11}{7})T^4_{\nu+\phi}$, we find
 the effective number of neutrinos
at matter radiation equality is
\begin{equation}
N_{\nu} = 3[1 + \frac{n'}{3 + \frac{11}{7}}]^{1/3}
\end{equation}

When the temperature drops to $T\sim m_{\phi}$ the number of
$\phi$'s get depleted via annihilation into standard model
neutrinos. In this case the neutrino temperature is
\begin{equation}
\frac{T_{\nu}}{T_{\gamma}} = (\frac{4}{11})^{1/3}[1 +  \frac{n' +
11/7}{3}]^{1/12}
\end{equation}
As an example if $n' = 3$ the effective number of neutrinos after
BBN is $ N_{\nu} = 4.08 $ and it is still constistent with CMB
data. Future CMB experiments like PLANCK \cite{Planck} and CMBpol
\cite{Polarization} will be able to improve the limit on $N_{\nu}$
and can provide a test of this model. The contribution of the
active neutrinos to the critical energy density is
\begin{equation}
\Omega_{\nu} = (\frac{53}{21})^{1/3}\frac{\sum{m_{\nu}}}{92h^2}
\label{eq1}\end{equation}
Using the upper recent bound on the neutrino energy density
\cite{Seljak} on finds
\begin{equation}
\sum{m_{\nu}} < 0.36\;eV
\end{equation}
We emphasize that this limit is only on the sum of the masses of the
active neutrinos since in our model the sterile neutrinos have decayed
away.

Incidentally, the same steps can be repeated for theories without
supersymmetry. In which case, we will assume that the $\phi$ field has
only a real part $\rho$ and an imaginary part $\chi$. The $\rho$ field
will have mass of order of the $\phi$-vev or about 100 keV whereas we will
assume the $\chi$ mass to be an eV. In this case, below $T\simeq m_e$,
first $\rho$ decay dumps into the $\nu+\nu_s+\chi$ system giving
\begin{eqnarray}
T_{\nu+\nu_s+\chi}~= 
~\left(\frac{n'+3+\frac{8}{7}}{n'+3+\frac{4}{7}}\right)^{1/3}
T_{\nu+\nu_s+\phi}
\end{eqnarray}
where $T_{\nu+\nu_s+\phi}\simeq
\left(\frac{3}{3+\frac{8}{7}+n'}\right)^{1/4}T^0_\nu$ due to
$\nu-\nu_s-\phi$ equilibrium.
Taking entropy conservation at the subsequent $\nu_s$ decay and
$\chi\chi\rightarrow \nu\nu$ as in the previous case, we get
\begin{eqnarray}
\frac{T_\nu}{T_\gamma}~=~\left(\frac{4}{11}\right)^{1/3} 
\left(\frac{3+\frac{8}{7}+n'}{3}\right)^{1/12}
\end{eqnarray}
This changes the coefficient in the formula in Eq.\ref{eq1} to
$\left(\frac{50}{21}\right)^{1/3}$ and changes on the limit on the sum of
active neutrino masses from 0.36 eV to 0.37 eV.

 Our scenario for sterile neutrinos has
also interesting
astrophysical implications. The first point is to look for any new
mechanism for energy loss from the supernova core via emission of $\nu_s$
or $\phi$. since $\nu-\nu_s$ mixing
arises from spontaneous symmetry breaking at scale $\ll$ MeV,
inside hot astrophysical environments such as a supernova, the
active and sterile neutrinos remain unmixed. As a result, there is no
energy loss via the emission of $\nu_s$. However, there could be energy
loss due to the processes $\nu\nu\rightarrow \phi\phi, \nu_s\nu_s$. The
rates for these
processes are estimated to be: $\sim 10^{-25}T\sim
(50~sec)^{-1}$. Comparing this with typical supernova explosion time
scale, we expect this energy loss mechanism not to be significant. Also
due to zero mixing between $\nu-\nu_s$, all supernova results based
on three active neutrinos\cite{dighe} remain unaffected. Only in the very
outer layers of the supernova explosion when the tempetature drops below
100 KeV, will these mixings become operative.

\section{Discussion and summary}
In summary, we have presented a mirror model for the sterile
neutrinos that can explain the LSND results and yet be consistent
with stringent constraints from big bang nucleosynthesis as well
as cosmic microwave background as well as structure formation
bounds on neutrino properties. We make predictions for the
effective neutrino number to which the next generation CMB
measurements are sensitive. An important requirement of this model
is that the reheat temperature after inflation must be less than a
TeV. The model has also other interesting properties discussed
earlier such as the mirror hydrogen being a dark
matter\cite{teplitz}, which remain unaffected by our modification.
Similarly, suggestions that the ultra high energy neutrinos could
be originating from topological defects in the mirror
sector\cite{vilenkin} remain unchanged by our extension.

\section*{Acknowledgments}
\vskip -.5cm This work is supported by National Science Foundation
Grant No. PHY-0354401.

\end{document}